\documentclass[
 ,final            % use final for the camera ready runs
  ]
  {aipproc}

\layoutstyle{6x9}

\begin{document}

\title[Accessing $\Delta$G with $\pi^{\pm}$ at PHENIX]{Accessing the Sign and Magnitude of $\Delta$G via high $p_{T}$, $A_{LL}^{\pi^{\pm}}$ in Polarized p+p Collisions at PHENIX}

\author{Astrid Morreale on behalf of the PHENIX Collaboraton}{
  address={University of California, Riverside, CA, 92521, USA}
  ,altaddress={http://www.phenix.bnl.gov,\\ Brookhaven National Laboratory, Bldg 510C, Upton, NY, 11973, USA} % additional visiting address
}

\begin{abstract}
The double-helicity asymmetries ($A_{LL}$) of $\pi^{+}$ and $\pi^{-}$ production in
polarized proton proton collisions are specially interesting probes of the gluon's polarization. Relative differences
among $A_{LL}$ of positive, neutral, and negative pions at high transverse momentum are sensitive to the sign
and of magnitude of $\Delta$G. Quark-gluon (qg) scattering starts to dominate mid-rapidity pion production at RHIC
at transverse momenta above $~$5GeV/c. In this kinematic region the favored and unfavored fragmentation functions
for each pion species are sensitive to both the gluon and the quark distributions, with different flavors having
different weights for each pion species. Charged pion asymmetry measurements will be an important component in future global
analyses, aiming to determine the gluon polarization over a wide range in x, the most recent results using polarized longitudinal data at $\sqrt{s}=$200GeV, are presented. 
\end{abstract}

\maketitle
%%%%%%%%%%%%%%%%%%%%%%%%%%%%%%%%%%%%%%%%%%%%
%% MAINMATTER
%%%%%%%%%%%%%%%%%%%%%%%%%%%%%%%%%%%%%%%%%%%%
\section{Introduction}
Measurements of $\pi^{\pm}$ $A_{LL}$ are known to be an important process in the $\Delta$G global analysis that aims to disentangle all of the partonic contributions to the proton spin puzzle.\cite{Bass2007}  With polarized proton-proton collisions and measurements of pions at PHENIX central rapidities, spin assymetries are valuable tools used to access information about the gluon's polarization. Under factorization, a differential cross section can be written as the convolution of a parton density function (pdf) and a
hard scattering process.This, along with universality of pdf's and fragmentation functions (FF, $D_{h}$,) allows for separation of
long and short distance(large momentum transfer interaction) contributions.  In practice, what is measured are the asymmetries: the ratio of the
polarized to unpolarized cross sections (Eq.\ref{eq1}). Asymmetries give an elegant way of accessing parton information by counting
observed particle yields in different helicity states of incident protons ($++, --$, versus, $+-, -+$) normalized by the polarization in
each beam ($P_{B,Y}$):
\newcommand{\sz}{\hspace*{-6pt}}
\begin{eqnarray}
 A^{\pi}_{LL} &\sz=\sz&  \frac{\displaystyle\sum_{a,b,c={\rm q}, \bar{\rm q}, {\rm g}} 
\Delta f_{a}\otimes \Delta f_{b} \otimes \Delta \hat{\sigma}\otimes D_{\pi/c}} 
{\displaystyle\sum_{a,b,c={\rm q}, \bar{\rm q}, {\rm g}} f_{a} 
\otimes f_{b} \otimes \hat{\sigma} 
\otimes D_{\pi/c}} = \frac{\sigma_{++} -\sigma_{+-}}{\sigma_{++}+\sigma_{+-}} \,, 
\nonumber\\
 A^{\pi}_{LL} &\sz=\sz& \frac{1}{P_{B}P_{Y}} \frac{N^{++}-RN^{+-}}{N^{++}+RN^{+-}}\,, 
\qquad\qquad  
R(luminosity) = \frac{L_{++}}{L_{+-}} \,.
\label{eq1}
\end{eqnarray}
\section{$\pi$ Mesons}
As pions form an isospin triplet, asymmetry measurements from all three pion species (for 5 $<$ $p_{T}$ $<$ 12 GeV/c) are particularly sensitive to the sign of $\Delta$G as qg interactions dominate pion production in this $p_{T}$ range. Neutral pions have been already measured and provide some constraints to the magnitude of the gluon's polarization.\cite{Boyle2007, Boyle} Charged pions are a model independent probe which are sensitive to both the sign and magnitude of $\Delta$G as preferential fragmentation of up quarks (u) to $\pi^{+}$, and down quarks(d) to $\pi^{-}$, leads to the dominance of u-g, and d-g contributions. This dominance of u or d combined with the different signs of their polarized distributions translates into asymmetry differences for the different species $\pi^{+}$ , $\pi^{0}$ and $\pi^{-}$ that depend on the sign of $\Delta$G. For example, a positive $\Delta$G could be indicated by an order of $\pi$ asymmetries, i.e: $A_{LL} (\pi^{+}) > A_{LL} (\pi^{0}) > A_{LL} (\pi^{-})$, and viceversa for a negative contribution. 
\subsection{PHENIX Detection of $\pi^{\pm}$}
An important characteristic of the PHENIX detector at RHIC his the fine-grained calorimetry-100 times finer than previous collider detectors. The resolution of the electromagnetic calorimeter (EMCal)is $\delta\eta*\delta\phi= 0.01*0.01$ \cite{Adcox2003}, allowing for high precision cluster identification. A high charged $p_{T}$ signal is selected by requiring an energy cluster in the EMCal in coincidence with the collision trigger. The charged particle which triggered the event is then associated to a high quality track in the drift chamber; only these charged triggered particles are then selected. The $\pi^{\pm}$ signal is selected with pion \v{C}erenkov light emision thresholds in the ring imaging \v{C}erenkov detector (RICH). Electromagnetic shower-shape cuts in the EMCal clusters aid in the purification of the sample. An additional $\frac{E}{p}$ cut is also used, where E is the energy deposited in the EMCal (one interaction length) and p is the momentum measured using the drift chambers. The main background for this measurement comes from electron conversions which proceed far from the collision vertex and will produce light in the \v{C}erenkov counter with high efficiency, albeit with very low energy. Background is removed by comparing a model of the charged track with the measured track signal. The background's vertex will show high deviation of the comparisong of model and track. The estimated remaining background in the sample is less than $2\%$ in the highest $p_{T}$ bin (7-10 GeV/c), and less than $1\%$ in the lower $p_{T}$ bins(5-7GeV/c) for both pion charges. 
\section{Results}
The data used in this measurement consist of 56,081(47,662) $\pi^{+}(\pi^{-})$ identified with the PHENIX detector using 2006 longitudinally polarized data. The approximate integrated luminosity analyzed was 5.8$pb^{-1}$ which is more than twice the luminosity and more than 3 times the figure of merit ($Luminosity*Polarization^{4}$) of the previous reported measurement \cite{Morreale2007}. The measured charge separated pion double spin assymetries are shown in Figure 1. The measurements are compared with previous charged pion results and with neutral pion assymetries, see figure 2. Figure 3 shows a comparison of these results to new GRSV parametrisations. These new parametrisations are calculated using the de Florian-Sassot-Stratmann (DSS) fragmentation functions ($D_{\pi^{\pm}}$) , which use the most updated $e^{\pm}$ and RHIC data. These calculations contain the best information on fragmentation functions current available \cite{Florian2007}, unlike the previously used KKP curves, also distinguish between positive and negative charges.[Fig. 3]
%%\paragraph{<A subsubsubsection>}
%%CITE \cite{Brown2000,BrownAustin:2000}. 
%%CITE 2\cite{Mittelbach/Schoepf:1990} CITE3 \cite{Wang} 
%%%%%%%%%%%%%%%%%%%%%%%%%%%%%%%%%%%%%%%%%%%%
%% Sample figure:
%%
%% The option [height=...] scales the picture to the given height,
%% without it it would be printed at its nominal size
%%%%%%%%%%%%%%%%%%%%%%%%%%%%%%%%%%%%%%%%%%%%

\begin{figure}
  \includegraphics[height=.25\textheight]{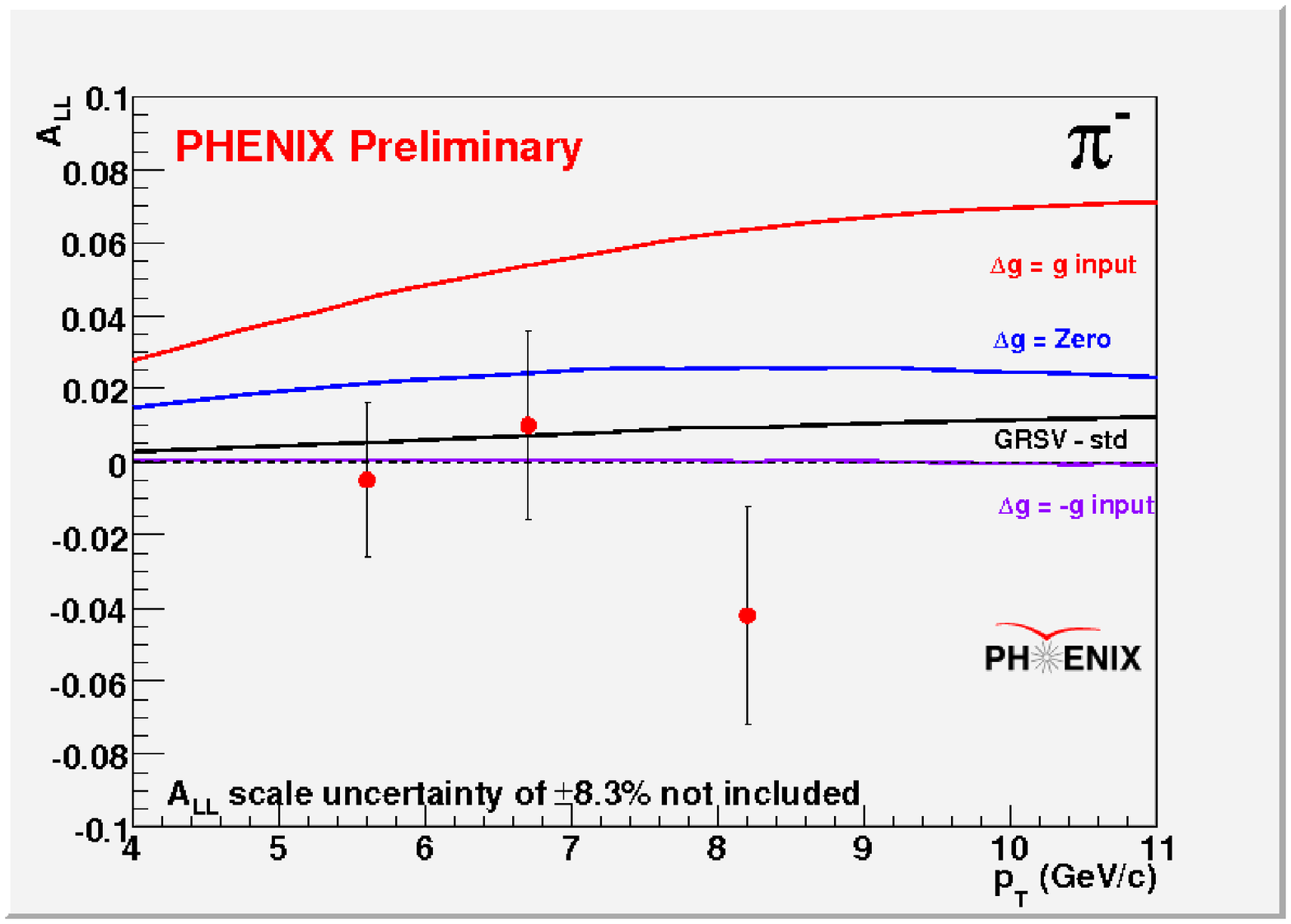}\includegraphics[height=.25\textheight]{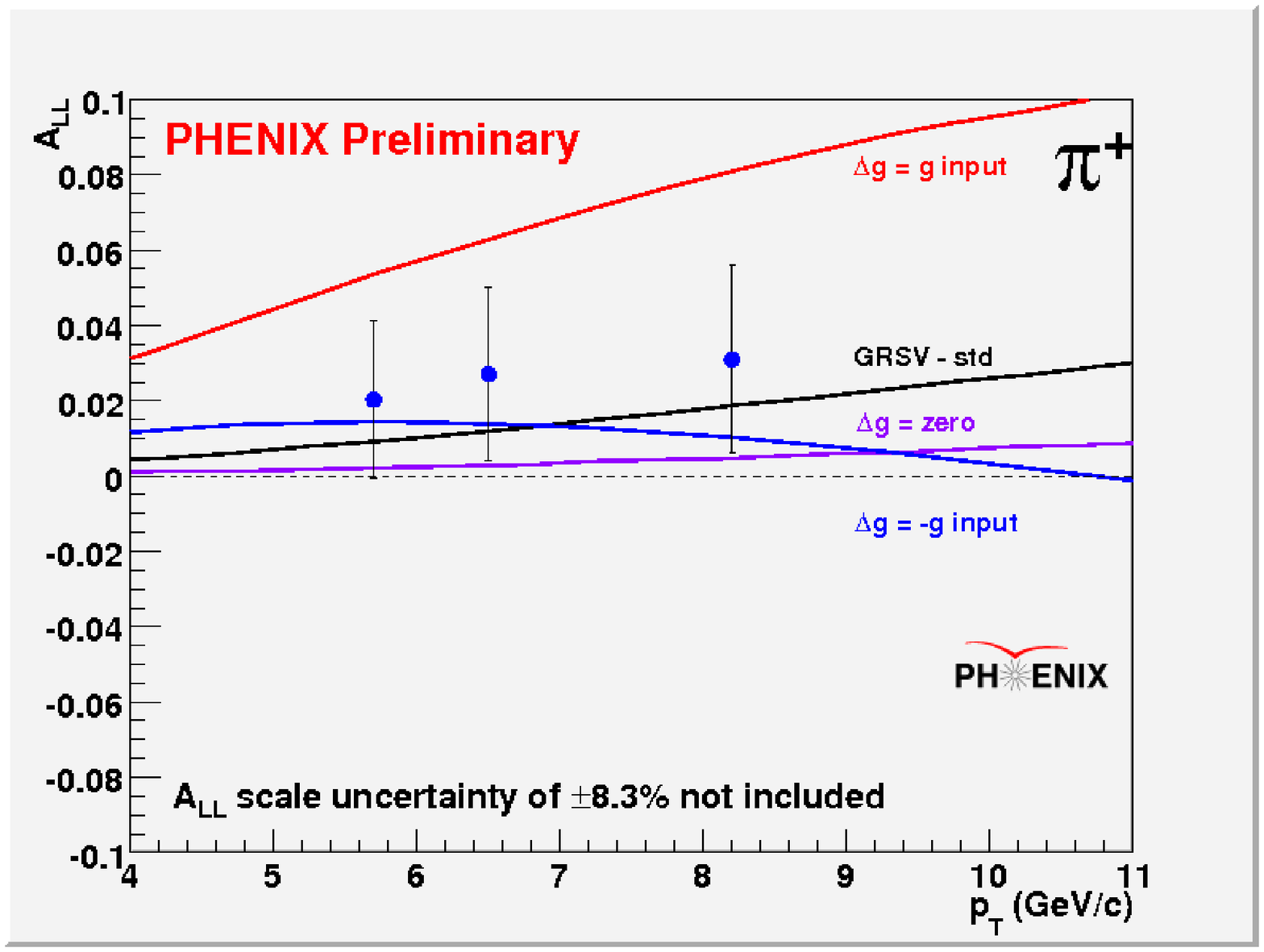}
  \caption{Measured of $A_{LL}$ of $\pi^{-}$(left) and $\pi^{+}$(right)}
\end{figure}
\begin{figure}
  \includegraphics[height=.25\textheight]{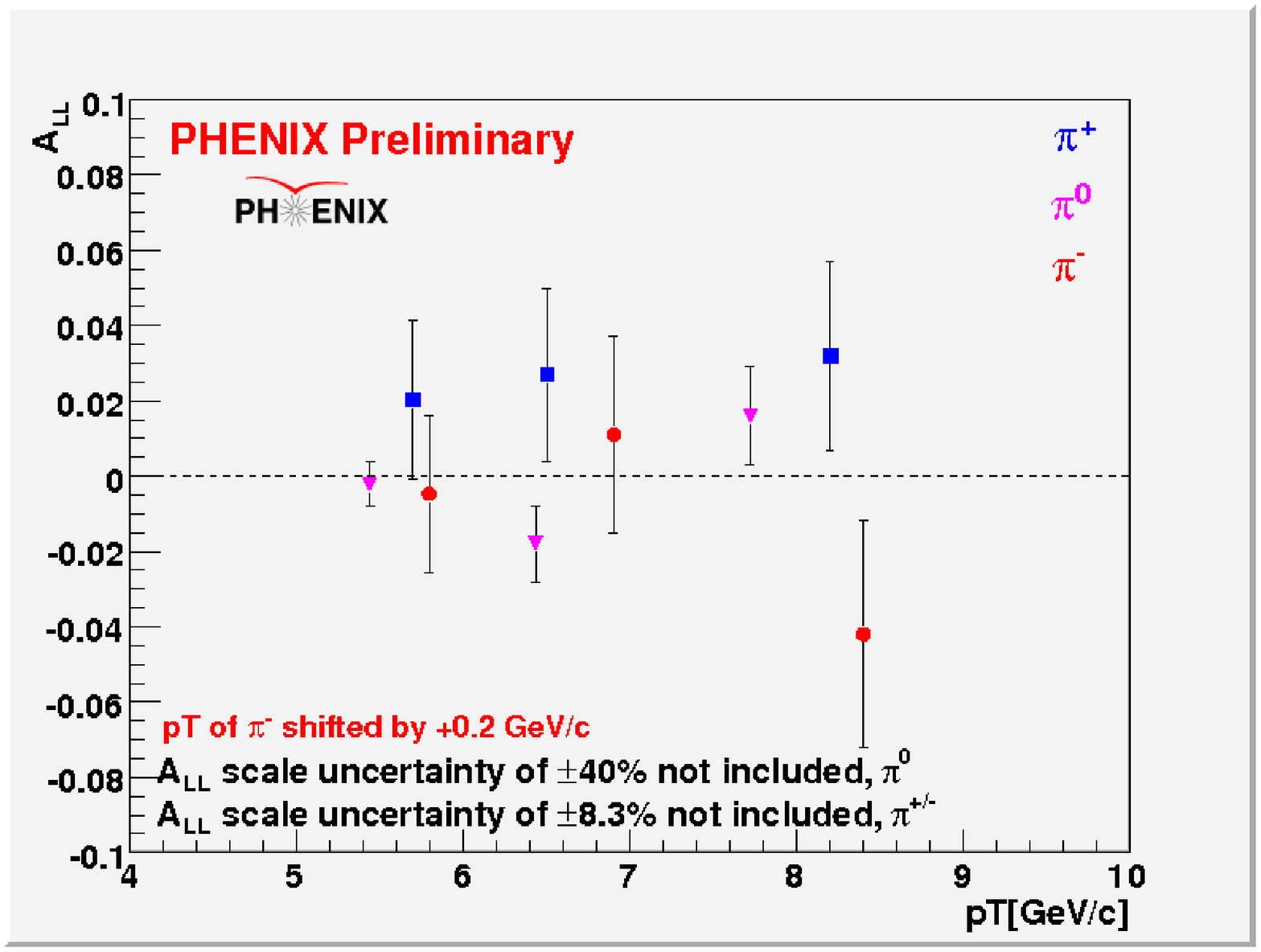}\includegraphics[height=.25\textheight]{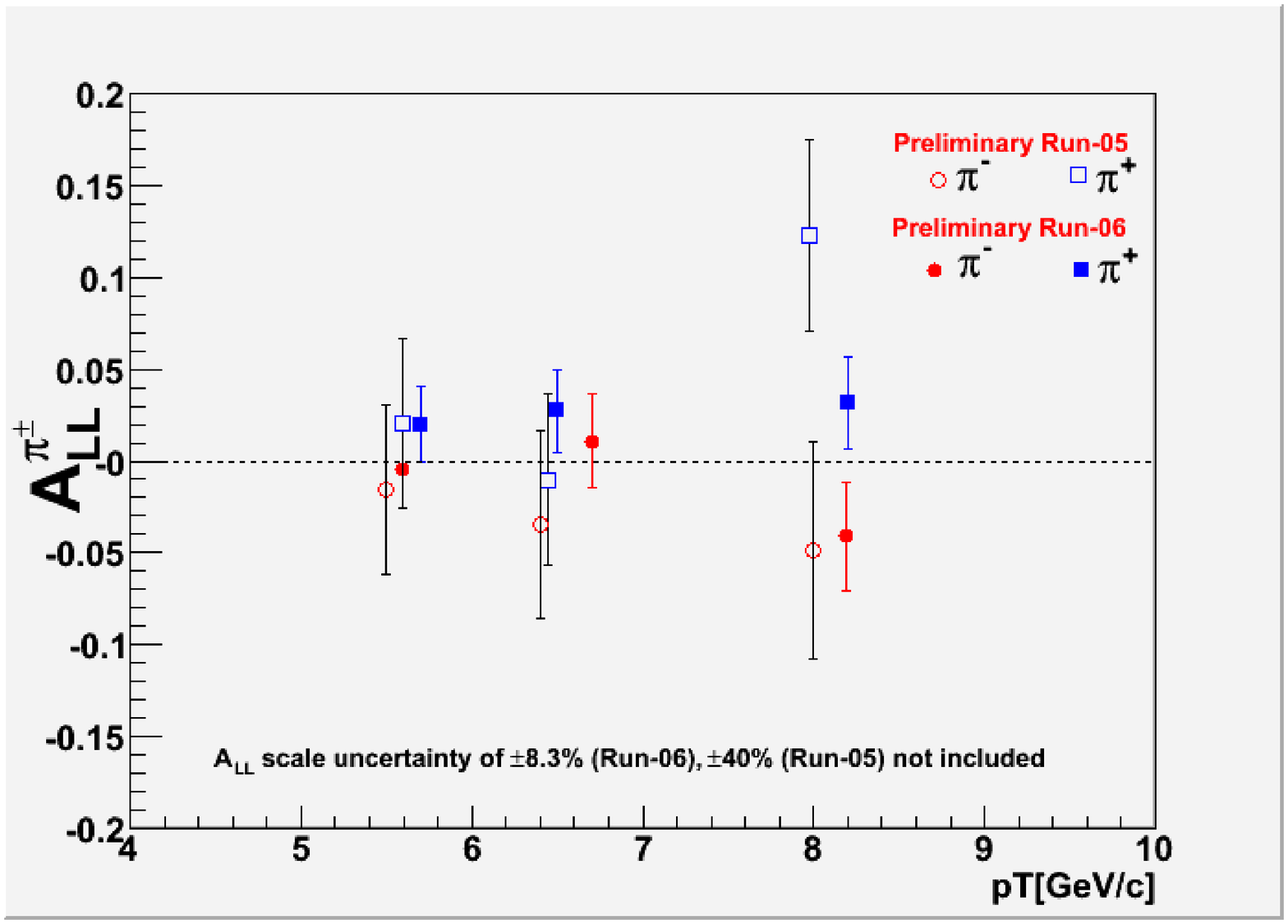}
  \caption{Comparison of charged and neutral assymetries at high pT(left), and previous reported measurements\cite{Morreale2007} }
\end{figure}
\begin{figure}
  \includegraphics[height=.25\textheight]{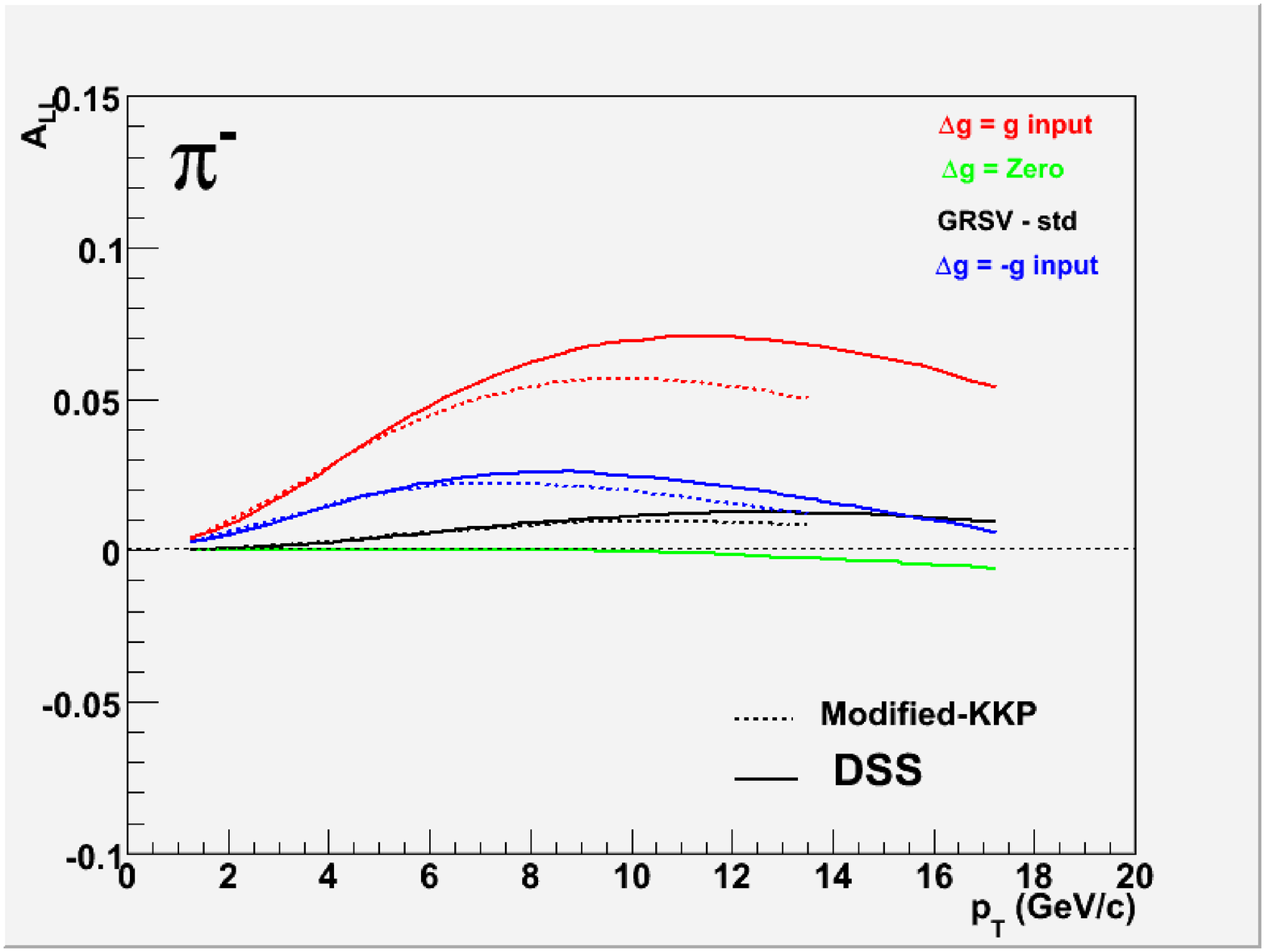}\includegraphics[height=.25\textheight]{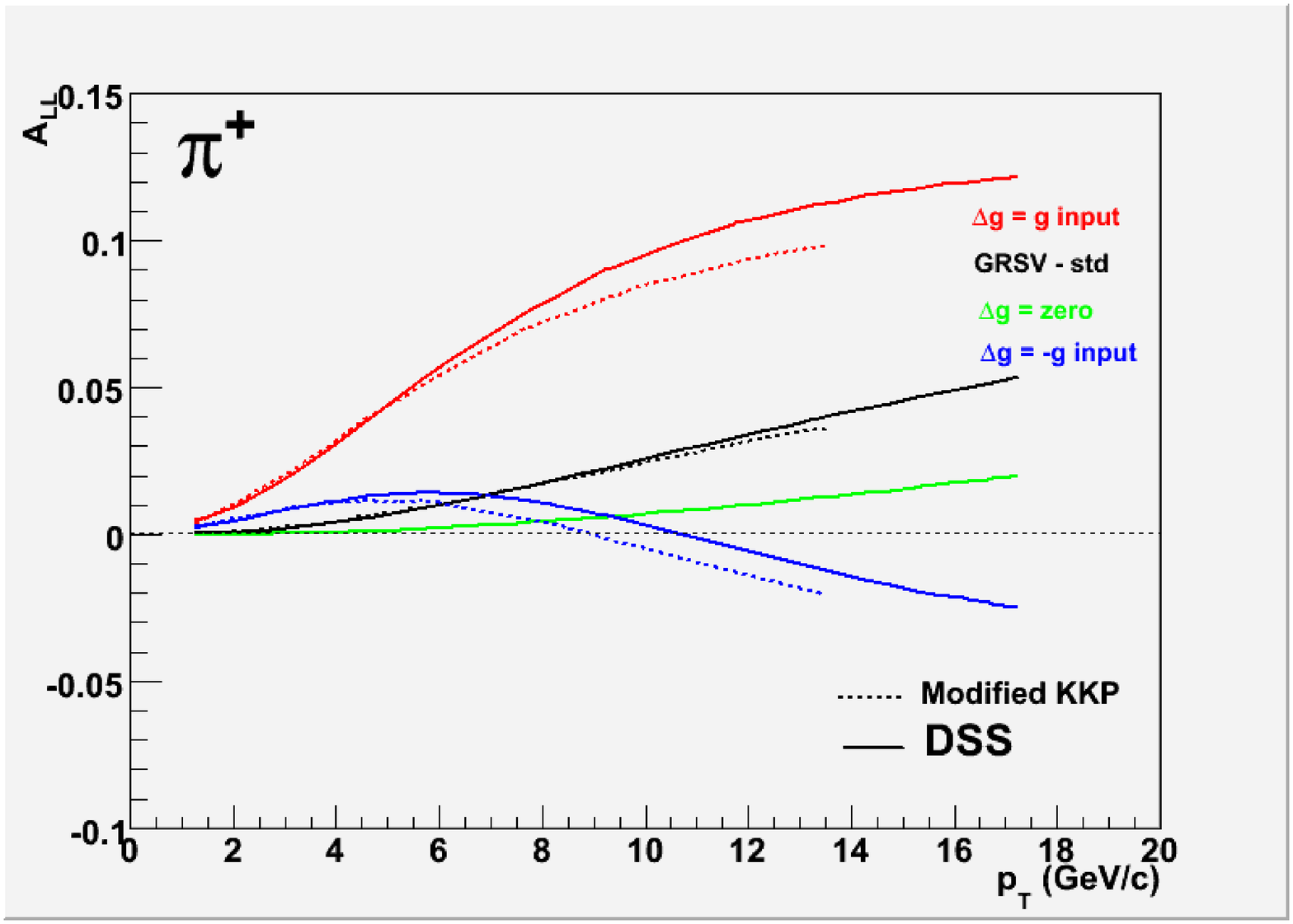}
  \caption{Comparison of GRSV curves using KKP and DSS fragmentation functions\cite{Florian2007}}
\end{figure}

\begin{figure}
  \includegraphics[height=.25\textheight]{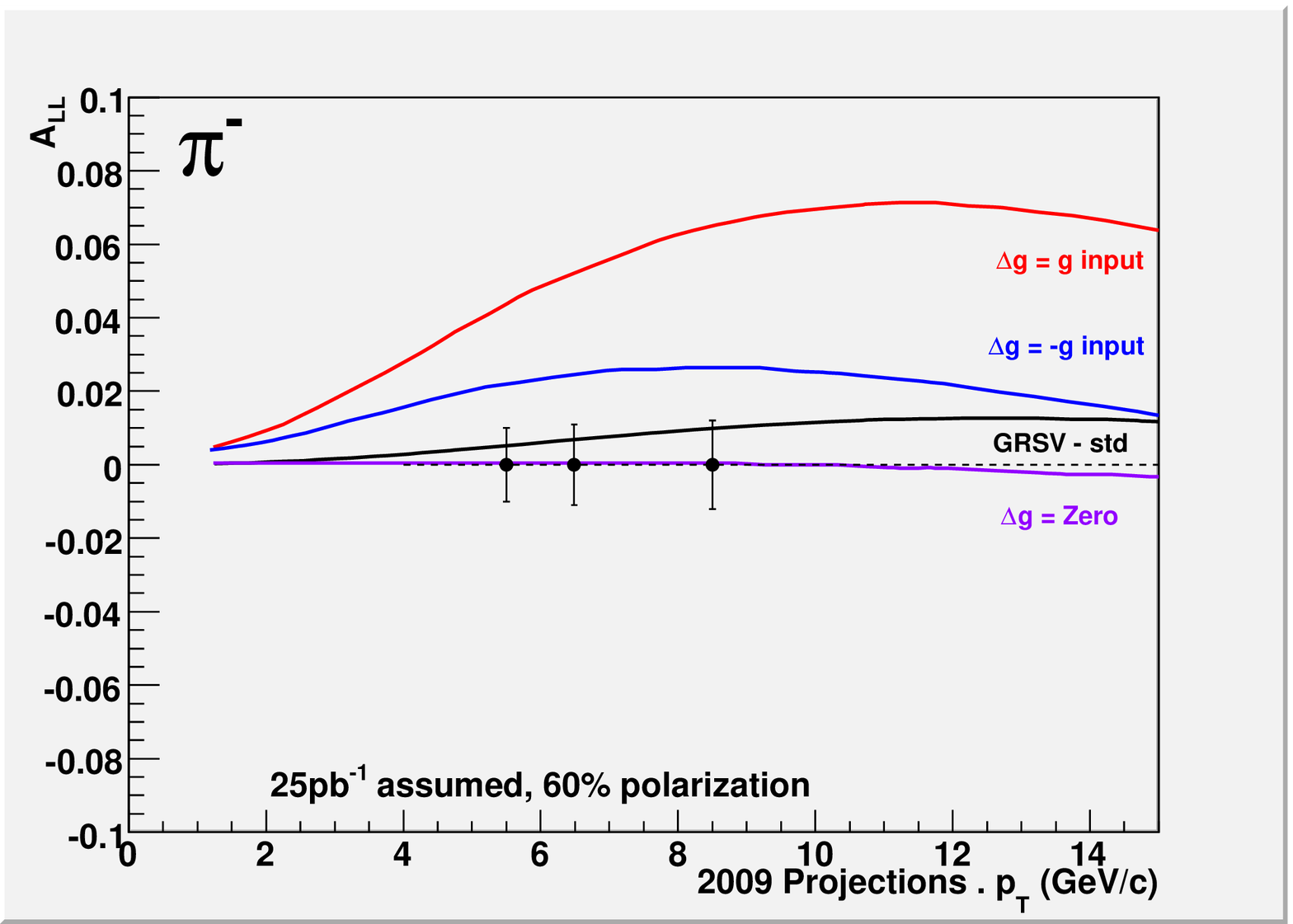}\includegraphics[height=.25\textheight]{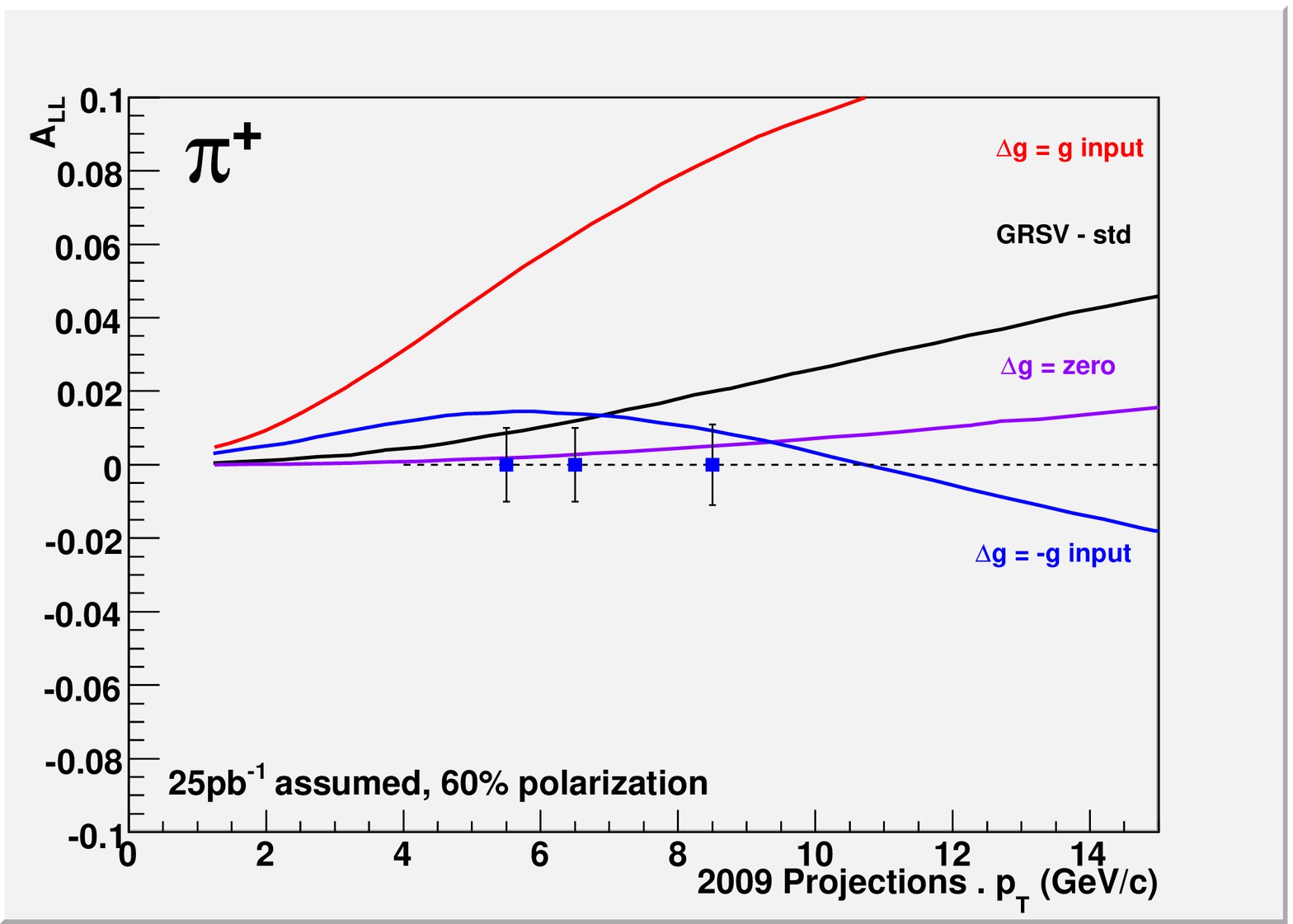}
  \caption{Charged pion sensitivity projections}
\end{figure}

%%\begin{description}
%%\item[Infandum]
 %%regina, iubes renovare dolorem, Troianas ut opes et lamentabile
 %%regnum cruerint Danai.
%%\item[Sed]
 %%si tantus amor casus cognoscere nostros et breviter Troiae supremum
 %%audire laborem, quamquam animus meminisse horret, luctuque refugit,
 %%incipiam.
%%\item[Lamentabile] regnum cruerint Danai; quaeque ipse miserrima vidi, et
%%quorum pars magna fui. Quis talia  fando Myrmidonum Dolopumve aut duri
%%miles Ulixi temperet a lacrimis?
%%\end{description}

\section{Conclusions}
The measured $\pi^{\pm}$ $A_{LL}$ (2006) have been compared with a new set of theoretical calculations. These for the first time use charge separated $\pi^{\pm}$ data for $D_{\pi^{\pm}}$ extraction. While the result has statistical improvement from the 2005
measurement, further statistics along with a cross section measurement would be useful in interpreting the results and comparing with pQCD predictions.
The projections of charged pion sensitivities for future RHIC longitudinally polarized runs are found in figure 4. The analyzing power of $\pi^{\pm}$ asymmetries are expected to be major players in future global analyses of $\Delta g$ and are thus an important part of the 200GeV longitudinal program at RHIC. 

\end{document}